\documentclass[preprint,12pt]{elsarticle}

\usepackage{amssymb,amsmath,amsthm,bm}

\biboptions{sort&compress}

\journal{Nuclear Physics B}

\begin{document}

\begin{frontmatter}



\title{Millicharged neutrino with anomalous magnetic moment in rotating magnetized matter}


\author[1,2]{Alexander I. Studenikin}
\ead{studenik@srd.sinp.msu.ru}
\author[1]{Ilya V. Tokarev}
\ead{tokarev.ilya.msu@gmail.com}

\address[1]{Department of Theoretical Physics, Faculty of Physics, Lomonosov Moscow State University, Moscow 119991, Russia}
\address[2]{Joint Institute for Nuclear Research, Dubna 141980, Moscow Region, Russia}

\begin{abstract}

New exact solutions of the modified Dirac equation describing a neutrino with nontrivial electromagnetic properties in extreme background conditions are obtained. Within the quasi-classical treatment the effective Lorentz force that describes the neutrino propagation in the magnetized rotating matter is introduced. We predict the effect of the spatial separation of different types of relativistic neutrinos and antineutrinos (different in flavors and energies) by the magnetized rotating matter of a star. Low energy neutrinos can be even trapped inside the star. We also predict two new phenomena: a new type of the neutrino electromagnetic radiation (termed ``Light of (milli)Charged Neutrino'', $LC\nu$) and a new mechanism of the star angular velocity shift due to neutrinos escaping the star (termed ``Neutrino Star Turning" mechanism, $\nu S T$). The possible impact of the $\nu S T$ mechanism on a supernova explosion yields a new astrophysical limit on the neutrino millicharge $q_{\nu}<1.3\times10^{-19}e_0$. In addition, the $\nu S T$ mechanism can be also used to explain the origin of pulsar ``anti-glitches'' and ordinary glitches as well.

\end{abstract}


\end{frontmatter}


\section{Introduction}

Neutrino electromagnetic properties are among the most intriguing and exciting problems in modern particle physics. Within the Standard Model in the limit of massless neutrinos the particle electromagnetic properties vanish. However, in different extensions of the Standard Model a massive neutrino has nontrivial electromagnetic properties (for a review of the neutrino electromagnetic properties see \cite{Giunti:2008ve,Giunti:2014ixa,Broggini:2012df}). That is why it is often claimed that neutrino electromagnetic properties open ``a window to the new physics" \cite{Studenikin:2008bd}.

It is a common knowledge that for a massive neutrino in the easiest generalization of the Standard Model with inclusion of the right-handed neutrino the neutrino magnetic moment is not zero and is proportional to the neutrino mass, $\mu_\nu \approx 3 \times 10^{-19}\mu_B(m_\nu/1~{\rm eV})$ \cite{Fujikawa:1980yx}. Much greater values are predicted in other various Standard Model generalizations (for details see \cite{Giunti:2008ve,Broggini:2012df}). The best experimental limit on the neutrino magnetic moment \cite{Beda:2012zz} is given by GEMMA experiment
\begin{equation}
\label{magn_moment_limit}
\mu_{\nu}\leq2.9\times10^{-11}\mu_B.
\end{equation}

It is usually believed  \cite{Bernstein:1963qh} that the neutrino electric charge is zero. This is often thought to be attributed to gauge invariance and anomaly cancellation constraints imposed in the Standard Model. In the Standard Model of $SU(2)_L \times U(1)_Y$ electroweak interactions it is possible to get \cite{Babu:1989ex, Foot:1990uf,Foot:1992ui} a general proof that neutrinos are electrically neutral. The electric charges of particles in this model are related to the $SU(2)_L$ and $U(1)_Y$ eigenvalues by the Gell-Mann-Nishijima relation $ Q_{st}=I_{3} + \frac{Y}{2}$.  In the Standard Model  without right-handed neutrinos $\nu_{R}$ the triangle anomalies cancellation constraints (the requirement of renormalizability) lead to certain relations among particles hypercharges $Y$, that are enough to fix all $Y$, so that hypercharges, and consequently electric charges, are quantized \cite{Foot:1990uf,Foot:1992ui}. The electric charge quantization provides the particles charges to be equal integral multiples of one third the electron charge that together with the present experimental limits on particles charges gives a final proof that neutrinos are electrically neutral.

However, if the neutrino has a mass, the statement that a neutrino electric charge is zero is not so evident as it meets the eye. The strict requirements for charge quantization may disappear in extensions of the standard  $SU(2)_L \times U(1)_Y$ electroweak interaction models if right-handed neutrinos $\nu_R$ with $Y\neq0$ are included. In this case the uniqueness of particles hypercharges $Y$ is lost (hypercharges are no more fixed) and in the absence of hypercharge quantization the electric charge gets ``dequantized'' \cite{Foot:1990uf,Foot:1992ui}. As a result, neutrinos may become electrically millicharged particles.

In general, the situation with charge quantization is different for Dirac and Majorana neutrinos. As it was shown in \cite{Babu:1989ex}, the charge dequantization for Dirac neutrinos occurs in the extended Standard Model  with right-handed neutrinos $\nu_R$ and also in a wide class of models that contain an explicit $U(1)$ symmetry. In such models an addition to the electric charge $Q_{st}$ that is determined by the barion $B$ and lepton $L$ numbers of a particle appears \cite{Foot:1992ui},
\begin{equation}
\label{Q}
Q=Q_{st}+(B-L)\varepsilon,
\end{equation}
where $\varepsilon$ is an arbitrary parameter of the theoretical model. Thus, the well known fundamental Gell-Mann-Nishijima relation between electric charge, isospin and hypercharge is modified accordingly. Eq. (\ref{Q}) for a neutrino charge yields
\begin{equation}
\label{varepsilon}
q_{\nu}=-q_0, \quad q_0=\varepsilon e_0,
\end{equation}
where the absolute value of the neutrino millicharge $q_0$ is expressed in units of the absolute value of the electron charge $e_0$ (for definiteness, we suppose a negative neutrino millicharge). Note, that the existence of the nonzero neutrino millicharge should not violate the charge conservation law and should lead to non-standard electric charges of quarks and leptons. On the contrary, if the neutrino is a Majorana particle, the arbitrariness of hypercharges in this kind of models is lost, leading to electric charge quantization and hence to neutrino neutrality \cite{Babu:1989ex}.

The most model-independent limit on the neutrino electric charge was derived from the absence of an anomalous dispersion of the SN 1987A neutrino signal \cite{Raffelt:1996wa,Barbiellini:1987zz} which led to $q_0\leq3\times10^{-17}e_0$. Note that the most severe experimental constraint on the neutrino electric charge $q_0\leq3\times10^{-21}e_0$ is obtained from the electric charge conservation in a neutron $\beta$-decay \cite{Marinelli:1983nd,Baumann:1988ue}.

In this paper we consider Dirac neutrinos with nontrivial electromagnetic properties in extreme background conditions, namely in the magnetized rotating matter. We develop the approach (the ``method of exact solutions") that implies the use of the exact solutions of the modified Dirac equation for a neutrino wave function in the presence of the background environment \cite{Studenikin:2004dx,Grigorev:2005sw,Grigoriev:2007zzc,Studenikin20066769,Studenikin2006289,Studenikin:2008qk,Balantsev:2010zw}. On the basis of the obtained solutions we predict new astrophysical effects and phenomena which are produced by the interactions of millicharged neutrinos with the background magnetized rotating matter.

In section II the new exact solution of the Dirac equation describing the millicharged neutrino with the anomalous magnetic moment in a magnetized matter is obtained. In section III we obtain another one new solution of the Dirac equation describing the massless millicharged neutrino in the magnetized rotating matter. Section IV is devoted to the quasi-classical interpretation of the obtained solutions. In particular, we predict the existence of the effective Lorentz force that describes neutrino motion in the magnetized rotating matter. The effective Lorentz force contains the ordinary electromagnetic Lorentz force and the additional part that comes from the weak interactions of neutrinos with the matter. On this basis we predict the effect of spatial separation of different types of neutrinos and antineutrinos (different in flavors and energies) propagating in the rotating magnetized matter. This effect can be very important for neutrino searches from the transient astrophysical sources with optical follow-up observations~\cite{Ageron:2012mz}.

In section V a new mechanism of the neutrino electromagnetic radiation in the rotating magnetized nonuniform matter due to the neutrino millicharge (``Light of (milli)Charged Neutrino'',~$LC\nu$) is considered. This electromagnetic radiation phenomenon exists even in the absence of electromagnetic fields and, therefore, is of different nature than the one of the cyclotron radiation of charged particles in magnetic fields. The $LC\nu$ mechanism is of interest for astrophysics in light of the recently reported hints of ultra-high energy neutrinos~$\sim$ PeV observed by IceCube~\cite{Abbasi:2012cu}.

In section VI we introduce a new mechanism of a star rotation frequency shift produced by the feedback of escaping neutrinos to the star matter (termed the ``Neutrino Star Turning" mechanism, $\nu S T$). Accounting only for weak interactions the change of the star rotation rate is significant only in case of a sufficiently large neutrino emission.

In section VII we consider the electromagnetic part of the $\nu S T$ mechanism. In order to avoid the contradiction between the possible impact of the $\nu S T$ mechanism on the rotation rate of a pulsar during its formation in a supernova explosion with the initial pulsar rotation rate we obtain a new astrophysical limit on the neutrino millicharge $q_{\nu}<1.3\times10^{-19}e_0$. The limit is stronger than many other constraints known in literature \cite{Raffelt:1996wa} and it is indeed the best limit from astrophysics (see~\cite{Giunti:2008ve,Giunti:2014ixa}).

In section VIII we briefly discuss pulsar glitches and propose to use the $\nu S T$ mechanism to explain the origin of both glitches and ``anti-glitches'' phenomena. In particular, we estimate the total number of neutrinos that is needed to explain the recently observed ``anti-glitch'' event~\cite{Archibald:2013kla}.


\section{Millicharged neutrinos with anomalous magnetic moment in magnetized neutron matter}

The millicharged neutrino with the nonzero magnetic moment moving in an external magnetic field and dense neutron matter is described by the modified Dirac equation \cite{Studenikin:2013yaa,Balantsev:2012ep, Balantsev:2013aya}
\begin{equation}
\label{dirac}
\left(\gamma_{\mu}P^{\mu}-\frac12\gamma_{\mu}(1+\gamma_5)f^{\mu}
-\frac{i}{2}\mu\sigma_{\mu\nu}F^{\mu\nu}-m\right)\Psi(x)=0,
\end{equation}
where $P^{\mu}=p^{\mu}+q_0A^{\mu}$ is a particle kinetic momentum ($q_0$ is given by Eq.~(\ref{varepsilon})), $F^{\mu\nu}=\partial^{\mu}A^{\nu}-\partial^{\nu}A^{\mu}$ with $A^{\mu}=(0,-\frac{yB}{2},\frac{xB}{2},0)$ describes the uniform magnetic and $\mu$ is a neutrino anomalous magnetic moment. Note that the nontrivial neutrino electromagnetic properties can lead to additional electromagnetic interactions of the neutrinos with the particles of matter that modify the potential $A^{\mu}\rightarrow A^{\mu}+\delta A^{\mu}$ in Eq.~(\ref{dirac}). However, in case of electrically neutral matter in linear approximation the fluctuations $\delta A^{\mu}$ can be safely neglected due to the presence of the strong background magnetic field given by $A^{\mu}$.

The matter potential $V_m=\frac12\gamma_{\mu}(1+\gamma_5)f^{\mu}$ comes from the weak interactions of the neutrino with the background matter. The explicit form of $f^{\mu}$ depends on the background particles densities, velocities and polarizations. In this section we consider the case of the unpolarized static neutron matter when $f^{\mu}=-Gn_n(1, 0, 0, 0)$, where $G=\frac{G_F}{\sqrt{2}}$ ($G_F$ is the Fermi constant) and $n_n$ is the neutron number density. Note that the case of the neutrino motion in the rotating magnetized matter will be considered in the next section.

Eq.~(\ref{dirac}) can be written in the Hamiltonian form $i\frac{\partial}{\partial t}\Psi(x)=H\Psi(x)$ with the Hamiltonian
\begin{equation}
\label{hamiltonian}
H=\gamma_0\bm{\gamma P}+\gamma_0m+\gamma_0\sigma_3\mu
B-(1+\gamma_5)\frac{Gn_n}{2}.
\end{equation}
The exact solution for the neutrino wave function can be obtained in the form
\begin{equation}
\label{solution}
\Psi(x)=
\sqrt{\frac{q_0B}{2\pi L}}e^{-i(p_ot-p_3z)}
\begin{pmatrix}
C_1\mathcal{L}_s^{l-1}(\frac{q_0B}{2}r^2)e^{i(l-1)\varphi} \\
iC_2\mathcal{L}_s^l(\frac{q_0B}{2}r^2)e^{il\varphi} \\
C_3\mathcal{L}_s^{l-1}(\frac{q_0B}{2}r^2)e^{i(l-1)\varphi} \\
iC_4\mathcal{L}_s^l(\frac{q_0B}{2}r^2)e^{il\varphi}
\end{pmatrix},
\end{equation}
where $\mathcal{L}_s^l(\frac{q_0B}{2}r^2)$ are the Laguerre functions ($N=l+s=0,1,2...$), here the polar coordinates are used (see \cite{Balantsev:2010zw}).

To advance in the exact solution of Eq.~(\ref{dirac}) one should determine the spin coefficients $C_i$. To this purpose we introduce a new type of the spin operator
\begin{equation}
\label{spin_op}
S=S_{tr}\cos\alpha+S_{long}\sin\alpha, \quad
\sin\alpha=\frac{Gn_n}{\sqrt{(Gn_n)^2+(2\mu B)^2}},
\end{equation}
which is the weighted superposition of the operators of longitudinal and transverse polarizations
\begin{equation}
\label{spin_long_tr}
S_{long}=\frac{\bm{\Sigma P}}{m}, \quad
S_{tr}=\Sigma_3+\frac{i}{m}
\begin{pmatrix}
0 & -\sigma_0 \\
\sigma_0 & 0 \\
\end{pmatrix}
[\bm\sigma\times\bm P]_3,
\end{equation}
where $\sigma_{\mu}$ are the Pauli matrixes. The angle $\alpha$ is determined by the neutron density $n_n$ and the magnetic field $B$. Note, that the same technique was used in \cite{Frolov:2007cm}. The spin operator~(\ref{spin_op}) commutes with the Hamiltonian~(\ref{hamiltonian}) and yields the spin integral of motion
\begin{equation}
\label{spin_spectrum}
\mathcal{S}=\frac{\zeta}{m}\sqrt{(m\cos\alpha+p_3\sin\alpha)^2+2Nq_0B}, \quad \zeta=\pm 1.
\end{equation}
Using the spin integral of motion~(\ref{spin_spectrum}) one can obtain the energy spectrum of the Hamiltonian~(\ref{hamiltonian})
\begin{equation}
\label{energy_spectrum}
p_0=-\frac{Gn_n}{2}+\varepsilon\sqrt{
p_3^2+2Nq_0B+m^2+\left(\frac{Gn_n}{2}\right)^2+(\mu B)^2+
2m\mathcal{S}\sqrt{\left(\frac{Gn_n}{2}\right)^2+(\mu B)^2}}
\end{equation}
where $\varepsilon=\pm1$. The spin coefficients are given by
\begin{equation}
\label{coefficients}
\begin{aligned}
C_1&=\frac12
\sqrt{1+\frac{m\cos\alpha+p_3\sin\alpha}{m\mathcal{S}}}
\sqrt{1-\sin(\alpha-\beta)},\\
C_2&=\frac12\delta_1\zeta
\sqrt{1-\frac{m\cos\alpha+p_3\sin\alpha}{m\mathcal{S}}}
\sqrt{1-\sin(\alpha+\beta)},\\
C_3&=\frac12\delta_2
\sqrt{1+\frac{m\cos\alpha+p_3\sin\alpha}{m\mathcal{S}}}
\sqrt{1+\sin(\alpha-\beta)},\\
C_4&=\frac12\delta_3\zeta
\sqrt{1-\frac{m\cos\alpha+p_3\sin\alpha}{m\mathcal{S}}}
\sqrt{1+\sin(\alpha+\beta)}.\\
\end{aligned}
\end{equation}
We use the notations
$\delta_1=\mathrm{sgn}[\sin\alpha-\cos\beta],\,
\delta_2=\mathrm{sgn}[\cos(\alpha-\beta)],\,\delta_3=\mathrm{sgn}[\cos\alpha+\sin\beta]$
and introduce a new angle $\beta$,
\begin{equation}
\cos\beta=\frac{p_3\cos\alpha-m\sin\alpha}{p_0+\frac{Gn_n}{2}}.
\end{equation}
It is easy to show that $\sin\beta$ introduced in Eq.~(\ref{coefficients}) is given by
\begin{equation}
\sin\beta=
\frac{\sqrt{\left(\frac{Gn_n}{2}\right)^2+\left(\mu B\right)^2}+m\mathcal{S}}{p_0+\frac{Gn_n}{2}}.
\end{equation}

Note that Eqs.~(\ref{solution}), (\ref{spin_spectrum}), (\ref{energy_spectrum}) and (\ref{coefficients}) represent the exact solution of the modified Dirac equation (\ref{dirac}) which describes the millicharged neutrino with the nonzero magnetic moment in the dense magnetized matter.

\section{Millicharged neutrinos in rotating magnetized neutron matter}

It is not an easy task to solve the Dirac equation in the most general form~(\ref{dirac}) in case of a moving matter. Therefore, to simplify the task we omit the neutrino mass and magnetic moment terms in Eq.~(\ref{dirac}) due to the obvious smallness of these values. The potential of the rotating matter  $V_m=\frac12\gamma_{\mu}(1+\gamma_5)f^{\mu}$ is described by $f^{\mu}=-Gn_n(1, -y\omega, x\omega, 0)$ (see also \cite{Balantsev:2010zw}). We consider the particular case of coincided directions of the magnetic field $\bm B$ and the matter rotation $\bm\omega$ ($\omega$ is an angular velocity of the matter rotation). Thus, the effective neutrino evolution Hamiltonian is given by
\begin{equation}
\label{hamiltonian_rot}
H=\gamma_0\bm{\gamma P}-(1+\gamma_5)(1+\gamma_0\gamma_1\omega y-\gamma_0\gamma_2\omega x)\frac{Gn_n}{2}.
\end{equation}

To describe neutrino spin properties in a moving matter we introduce a new type of the spin operator
\begin{equation}
\label{spin_oper_rot}
S=\frac{\bm{\Sigma P}}{m}-\left(\gamma_0\bm\gamma\bm v-\bm\Sigma\bm v\right)\frac{Gn_n}{2m},
\end{equation}
which consists of the operator of the longitudinal polarization $S_{long}$ (see Eq.(\ref{spin_long_tr})) and of the additional term that accounts for the matter motion. Using the spin operator~(\ref{spin_oper_rot}) we have obtained the neutrino energy spectrum
\begin{equation}
\label{energy_rot}
p_0=-\frac{Gn_n}{2}+\varepsilon\zeta\eta\left(m\mathcal{S}+\frac{Gn_n}{2}\right), \quad
\end{equation}
where $m\mathcal{S}=\zeta\sqrt{p_3^2+2Nq\mathcal{B}}$ is the spin integral of motion and $q\mathcal{B}=q_0B+(1-\varepsilon\zeta\eta)Gn_n\omega$. The physical interpretation of new values $q$ and $\mathcal{B}$ will be done in the next section within the quasi-classical treatment of the obtained solutions. For the neutrino wave function we get
\begin{equation}
\label{wave_func_rot}
\Psi(x)=
\frac{e^{-i(p_ot-p_3z)}}{2}\sqrt{\frac{q\mathcal{B}}{\pi L}}
\begin{pmatrix}
\frac{1-\varepsilon\zeta\eta}{2}\psi\\
\frac{1+\varepsilon\zeta\eta}{2}\psi
\end{pmatrix},
\end{equation}
where
\begin{equation}
\label{spinor_rot}
\psi=\begin{pmatrix}
\sqrt{1+\frac{p_3}{m\mathcal{S}}}\mathcal{L}_s^{l-1}\left(\frac{q\mathcal{B}}{2}r^2\right)e^{i(l-1)\varphi}\\
i\zeta\sqrt{1-\frac{p_3}{m\mathcal{S}}}C_2\mathcal{L}_s^l\left(\frac{q\mathcal{B}}{2}r^2\right)e^{il\varphi}
\end{pmatrix}.
\end{equation}

The new solutions (\ref{energy_rot}), (\ref{wave_func_rot})  and (\ref{spinor_rot}) describe the millicharged neutrino in the rotating magnetized neutron matter. The solutions depend not only on the energy sign ($\varepsilon=\pm1$) and the spin number ($\zeta=\pm1$) but also on the additional parameter $\eta=sign\left(1+\frac{Gn_n}{2m\mathcal{S}}\right)$ which appears in case of the massless neutrino. Note that for relativistic neutrinos $\eta=+1$.

From the structure of the wave function~(\ref{wave_func_rot}) it follows that the left-handed neutrino chiral state $\Psi_{L}\equiv\frac12(1+\gamma_5)\Psi$ corresponds to the negative neutrino spirality $\zeta=-1$ while the right-handed neutrino $\Psi_{R}\equiv\frac12(1-\gamma_5)\Psi$ corresponds to the positive spirality $\zeta=+1$. It is also obvious that the right-handed neutrino does not contain the matter terms in the obtained solutions and is attributed to a sterile neutrino.

From Eq.~(\ref{energy_rot}) the energy spectrum of the active neutrinos (left-handed neutrinos and right-handed antineutrinos) is just straightforward
\begin{equation}
\label{energy_nu_rot}
p_0=\sqrt{p_3^2+2N(q_0B+2Gn_n\omega)}-Gn_n.
\end{equation}
that is quantized due to both weak and electromagnetic interactions and represents the modified Landau levels of the millicharged neutrino in the rotating magnetized matter (see also \cite{Grigoriev:2007zzc,Studenikin20066769,Studenikin2006289,Studenikin:2008qk,Balantsev:2010zw,Studenikin:2013yaa,Balantsev:2012ep, Balantsev:2013aya}).

\section{Quasi-classical treatment}

Using the explicit form of the obtained solution (\ref{energy_rot}), (\ref{wave_func_rot})  and (\ref{spinor_rot}) one can calculate the root mean square radii of quasi-classical circular neutrino orbits ($p_3=0$, $N\gg1$)
\begin{equation}
\label{radius}
R^{2}=\int \Psi^{\dagger}\,r^{2}\,\Psi
\,d\textbf{r}=\frac{2N}{q_0B+2Gn_n\omega}.
\end{equation}
It is also possible to express the orbit radius in the completely classical form
\begin{equation}
\label{Omega}
R=\Omega^{-1}, \quad  \Omega=\omega_m+\omega_c,
\end{equation}
where we introduce the effective rotation frequency $\Omega$ which is determined by the cyclotron frequency
\begin{equation}
\label{omega_c_matter}
\omega_c=\frac{q_0B}{p_0+Gn_n},
\end{equation}
and the matter induced frequency
\begin{equation}
\label{omega_m_matter}
\omega_m=\frac{2Gn_n}{p_0+Gn_n}\omega.
\end{equation}

The radius~(\ref{radius}) is defined by the neutrino energy $p_0$ and by the background environment (namely, the neutron number density $n_n$, the angular velocity $\omega$ and the magnetic field $B$). In case of rather high angular matter velocity and strong magnetic field the radii of the quasi-classical orbits for low energy neutrinos become smaller than the size of the matter source. Thus, we predict the effect of low energy neutrinos trapping inside the rotating magnetized matter. For instance, low energy neutrinos can be trapped inside neutron stars or accretion disks of black holes.

The obtained quasi-classical neutrino circular orbits can be explained as a result of the action of the generalized effective Lorentz force (see also \cite{Studenikin:2008qk}). The force accounts for weak and electromagnetic interactions of the active neutrinos (left-handed neutrinos and right-handed antineutrinos) with the rotating nonuniform magnetized matter. The effective Lorentz force has the following form
\begin{equation}
\label{F_eff}
\bm F=q\bm{\mathcal{E}}+q
\left[\bm\beta\times\bm{\mathcal{B}}\right],
\end{equation}
where
\begin{equation}
\label{E_B_eff}
\begin{aligned}
q\bm{\mathcal{E}}&=q_m\bm{E}_m,\\
q\bm{\mathcal{B}}&=-(q_mB_m+q_0B)\bm{e}_z,\\
\end{aligned}
\end{equation}
$\bm\beta$ is the neutrino speed and $\bm{e}_z$ is a unit vector in the direction of the magnetic field and matter rotation.  The matter induced ``charge'' $q_m$, ``electric'' $\bm{B}_m$ and ``magnetic'' $\bm{E}_m$ fields are given by
\begin{equation}
\label{q_E_B_matter}
q_m=-G, \quad
\bm{E}_m=-\bm{\nabla}n_n, \quad
\bm{B}_m=-2n_n\bm\omega.
\end{equation}
The magnetic $\bm{B}$ field reproduces the ordinary electrodynamic Lorentz force and the matter induced components is due to $\bm{E}_m$ and $\bm{B}_m$. Note that in case of sterile neutrinos the effective Lorentz force (\ref{F_eff}) reduces to the ordinary electrodynamic Lorentz force.

The effective electric field $\bm{\mathcal{E}}$ is produced by the gradient of the matter number density. The matter induced charges of neutrinos and antineutrinos are of opposite signs, $q_m^{\nu}=-G$ and  $q_m^{\bar{\nu}}=G$. Therefore, neutrinos that propagate inside the neutron matter with decreasing density are decelerated while antineutrinos are accelerated. Note that the possibility of particle acceleration due to the gradient of the matter number density was also discussed in~\cite{Studenikin20066769,Studenikin2006289,Studenikin:2008qk,Balantsev:2010zw, OliveiraeSilva:2000ns,Loeb:1989nb}.

The effective magnetic field $\bm{\mathcal{B}}$ contains the magnetic field $\bm{B}$ itself and the additional term $\bm{B}_m=-2n_n\bm\omega$ originating by the matter rotation. As it follows from Eq.~(\ref{F_eff}), the  effective magnetic force is orthogonal to the neutrino speed $\bm\beta$ and, thus, bounds both neutrinos and antineutrinos.

The discussed above Lorentz force can have interesting consequences in astrophysics. In particular, the force reasonably disturbs the trajectory of neutrinos propagating inside the rotating magnetized matter. We predict a new effect of the spatial separation of different types of neutrinos (neutrinos and antineutrinos and/or different flavor neutrinos and/or neutrinos with different energies) propagating inside the matter.

Recently, attempts to find neutrino signals in the correlation with observed light signals from astrophysical transient sources (including gamma-ray bursts, core collapse supernovae and active galactic nuclei) give no results \cite{Ageron:2012mz}. The new effect of the neutrino deflection in the magnetized rotating matter can explain the absence of the neutrino signals in the corresponding experiments \cite{Ageron:2012mz}. Indeed, during the propagation of the initially collimated beams of neutrinos and photons the neutrinos are declined by the force (\ref{F_eff}) while the photons are not. Naturally, for relativistic neutrinos the deviation angle is very small,
\begin{equation}
\Delta\phi \simeq R_{S}\Omega\sin\theta,
\end{equation}
where $\theta$ is the azimuthal angle of the neutrino motion and $R_S$ characterizes the size of the source. However, due to immense distances to the astrophysical sources $L$ the deflection of the neutrino beam is huge,
\begin{equation}
\label{deflection}
\Delta L\simeq\Delta\phi L,
\end{equation}

If we account only for the weak interactions and consider the source distance $L\sim1$ kpc the deflection~(\ref{deflection}) is about the distance between the Earth and the Sun (for $\sin\theta=1$). Therefore, a neutrino beam can be reasonably deflected by the effective Lorentz force and cannot accompany the light signal in terrestrial observations even in the case of their initially coinciding directions. Thus, the effect discussed above can explain the ANTARES experimental results \cite{Ageron:2012mz}.

\section{Light of millicharged neutrinos}

The effective Lorentz force~(\ref{F_eff}) should lead to the neutrino acceleration and the corresponding emission of the electromagnetic radiation due the neutrino electric millicharge. In the most general case of the neutrino motion in nonuniform magnetized rotating matter the effective Lorentz force originates the neutrino acceleration
\begin{equation}
\label{acceleration}
\bm{a}=\frac1m\left(G\bm\nabla n_n+
(q_0B+2Gn_n\omega)
\left[\bm\beta\times\bm{e}_z\right]\right).
\end{equation}
The radiation power of the induced electromagnetic radiation is given by \cite{Schwinger:1949ym}
\begin{equation}
\label{radiation_power}
I_{LC\nu}=\frac{2q_0^2}{3}
\left(\frac{\dot{\bm\beta}^2}{(1-\bm\beta^2)^2}+
\frac{(\bm\beta\dot{\bm\beta})^2}{(1-\bm\beta^2)^3}\right).
\end{equation}
We term the considered mechanism of the neutrino electromagnetic radiation due to the neutrino millicharge, that can be emitted in the presence of the nonuniform rotating matter and electromagnetic fields, the ``Light of (milli)Charged Neutrino'' ($LC\nu$). It should be stressed, that the phenomenon exists even in the absence of the electromagnetic fields, when the acceleration~(\ref{acceleration}) is produced only due to the weak interactions of neutrinos with background particles (see also \cite{Studenikin:2008qk}). So that the proposed new mechanism is of different nature than the one of the cyclotron radiation of charged particles in magnetic fields.

The $LC\nu$ mechanism would manifest itself during the neutrino propagation from the central part of a rotating neutron star outwards through the crust. The gradient of the matter density (the density variation along the neutrino path) gives the following contribution to the $LC\nu$ radiation power (see Eq.~(\ref{acceleration}))
\begin{equation}
\label{rad_power_LC_line}
I_{LC\nu}=\frac{2q_0^2}{3m^2}(G\bm\nabla n_n)^2,
\end{equation}
and the effect of the matter rotation  yields
\begin{equation}
\label{rad_power_LC_circle}
I_{LC\nu}=\frac{2q_0^2}{3m^2}(q_0B+2Gn_n\omega)^2\gamma^2,
\end{equation}
where $\gamma=(1-\bm \beta^2)^{-1/2}$. The numerical estimations, that account for the $LC\nu$ power for the present limits on the neutrino millicharge and for a realistic gradient of a neutron star matter density $|G\bm\nabla n_n|\sim1 \text{eV}/1 \text{km}$ and the rotation frequency $\omega\sim2\pi\times10^3$ s$^{-1}$, show that the role of the $LC\nu$ in the explosion energetics is negligible in respect to the total energy of the collapse. However, as it was shown before \cite{Oraevsky:1994wb,Nieves:2003kw,Duan:2004nc}, in the presence of a dense plasma the induced neutrino effective electric charge can be reasonably large. In addition, the phenomenon is of interest for astrophysics in light of the recently reported hints of ultra-high energy neutrinos~$\sim$ PeV observed by IceCube~\cite{Abbasi:2012cu}.

\section{Millicharged neutrinos as the star rotation engine}

Consider the propagation of the neutrinos inside a neutron star on the basis of the obtained solutions for neutrinos in the rotating magnetized matter. During the propagation the effective Lorentz force (\ref{F_eff}) disturbs the neutrino trajectories. Obviously, there is also the feedback of the neutrinos on the star. Therefore, our prediction is as follows: neutrinos that propagate inside a star should effect the star rotation. This new effect we have termed the ``Neutrino Star Turning" ($\nu S T$) mechanism.

The idea of star angular momentum losses due to the neutrino emission was proposed for the first time in \cite{Mikaelian:1976ce,Epstein:1978md}. In these and other subsequent papers only the slow-down of the star rotation is considered. In our paper we develop a new approach to the description of neutrino propagation inside a rotating star on the basis of the introduced $\nu S T$ mechanism that can lead to both acceleration and deceleration of the star rotation.

To estimate the efficiency of the predicted $\nu S T$ mechanism we consider the impact of the escaping neutrinos on the star rotation. A single neutrino propagating with the azimuthal angle $\theta$ generates a feedback force with the projection onto the rotation plane given by
\begin{equation}\label{F}
F=(q_0B+2Gn_n\omega)\sin\theta.
\end{equation}
The value of the corresponding torque is just straightforward,
\begin{equation}\label{M_0}
M_{0}(t)=\sqrt{1-\frac{r^{2}(t)\Omega^2\sin^2\theta}{4}}Fr(t)\sin\theta,
\end{equation}
where $\bm r(t)$ is the neutrino radius vector inside the star and $\Omega$ is given by Eq.~(\ref{Omega}). Accounting for the contribution from all emitted neutrinos $N_{\nu}$ one can obtain the total torque
\begin{equation}\label{M}
M(t)=\frac {N_{\nu}}{4\pi}\int M_{0}(t) \sin \theta d\theta d\varphi,
\end{equation}
that determines the rate of change of the star angular momentum, $M(t)=I_S\frac{d\omega}{dt}$, where $I_{S}=\frac{2}{5}M_{S}R^3_{S}$ is the moment of inertia of the star with mass $M_{S}$ and radius $R_S$. In case of relativistic neutrinos we obtain the shift of the star angular velocity in the following form
\begin{equation}
\label{delta_omega}
|\triangle\omega|=\frac{5N_{\nu}}{6M_{S}}(q_0B+2Gn_n\omega_0),
\end{equation}
where $\triangle\omega=\omega-\omega_0$ ($\omega_0$ is an angular velocity of the star before the neutrino emission).

Recall that the matter induced charges (\ref{q_E_B_matter}) of neutrinos and antineutrinos are of opposite signs, $q_m^{\nu}=-G$ and  $q_m^{\bar{\nu}}=G$. Therefore, the weak interactions of neutrinos with the star matter spin down the star rotation ($\triangle\omega<0$) while for antineutrinos the rotation is spined up ($\triangle\omega>0$). The electromagnetic interaction of the negative millicharged neutrinos and positive millicharged antineutrinos with the star magnetic field reinforces these effects. The impacts of the weak and electromagnetic interactions on the $\nu ST$ mechanism are defined by the matter induced and cyclotron frequencies accordingly (see Eqs.(\ref{omega_c_matter}) and (\ref{omega_m_matter})).

To estimate the impact of the weak interactions on the $\nu ST$ mechanism one should set $q_0=0$ in Eq.~(\ref{delta_omega}). In this case the relative shift of angular velocity is given by
\begin{equation}
\label{delta_omega_rel}
\frac{|\triangle\omega|}{\omega_0}=4\times10^{-66}N_{\nu}
\left(\frac{1.4M_{\odot}}{M_{S}}\right)
\left(\frac{\rho_n}{10^{14}\textrm{g}/\textrm{cm}^3}\right),
\end{equation}
where $M_{\odot}$ is the solar mass and $\rho_n$ is the density of the neutron matter. From the obtained estimation it follows that the impact of the weak interaction on the pulsar angular velocity is of interest only in case of sufficiently large neutrino emission. However, the consideration of the electromagnetic interaction can significantly reinforce the discussed effect.

\section{New astrophysical limit on the neutrino millicharge}

One of the most strongest astrophysical neutrino sources is a supernova explosion with total $N_{\nu} \sim 10^{58}$ of emitted neutrinos \cite{Hirata:1987hu}. Therefore, the impact of the $\nu ST$ mechanism on the pulsar rotation rate during the formation of the pulsar in the supernova explosion is of particular interest. In case of a nonzero neutrino millicharge the $\nu ST$ mechanism is originated dominantly by the electromagnetic interactions of the neutrinos with the magnetic field of the pulsar since $\omega_c\gg\omega_m$. From Eq.~(\ref{delta_omega}) we obtain the relative frequency shift of a born pulsar due to the $\nu ST$ mechanism
\begin{equation}
\label{delta_omega_nonzero_charge}
\frac{|\triangle\omega|}{\omega_0}=7.6\varepsilon\times
10^{18}\left(\frac{P_0}{10\text{ s}}\right)
\left(\frac{N_{\nu}}{10^{58}}\right)
\left(\frac{1.4  M_{\odot}}{M_{S}}\right)
\left(\frac{B}{10^{14} G}\right),
\end{equation}
where $P_0$ is a pulsar initial spin period and $\varepsilon$ is the neutrino electric millicharge in units of the electron charge (see Eq.~(\ref{varepsilon})).

The current pulsar timing observations \cite{Rosen:2012ty} show that the present-day rotation periods are up to $10$ s. The rotation during the life of a pulsar spins down due to several various mechanisms and dominantly due to a magnetic dipole braking. However, all of the estimations of feasible initial pulsar spin periods $P_0$ give the values that are very close to the present observed periods. Therefore, the estimation~(\ref{delta_omega_nonzero_charge}) is given for $P_0=10$ s. Note that the estimation can be strengthen in case of superluminous supernovae recently discovered~\cite{Howell:2013qua}.

The possible existence of a nonzero neutrino millicharge should not significantly change the rotation rate of a born pulsar. From the straightforward demand $|\triangle\omega| < \omega_0$ and Eq.~(\ref{delta_omega_nonzero_charge}) we obtain a new limit on the neutrino millicharge
\begin{equation}
\label{bound_q_nu}
q_0<1.3\times10^{-19} e_0.
\end{equation}
That is, in fact, one of the most severe astrophysical limits on the neutrino millicharge \cite{Giunti:2008ve,Giunti:2014ixa}. Note that in \cite{Barbiellini:1987zz} the limit on the neutrino millicharge $\sim10^{-15} - 10^{-17} e_0$ was obtained from a quite different approach based on the consideration of the millicharged neutrino flux from SN1987A propagating in the galactic magnetic field outside the star.

\section{The $\nu ST$ mechanism and pulsar glitches}

The obtained result for the frequency shift~(\ref{delta_omega}) recalls a very intriguing phenomenon that occurs during a life of pulsars. Pulsar timing observations also show the effects of the sporadic sudden increase of the rotation frequency (the pulsar glitches). The mechanism of this phenomenon has been explained in many papers \cite{Anderson:1975zze}. According to the recent observations \cite{Rosen:2012ty}, the relative shift of the rotation frequencies is ranged between $10^{-10}$ and $10^{-5}$ with peaks at approximately $10^{-9}$ and $10^{-6}$.

Very recently the observation of the ``anti-glitch'' phenomenon, that is a sudden decrease of the rotation frequency, has been reported \cite{Archibald:2013kla}. The observation of the ``anti-glitch'' requires reexamination of the nature of the pulsars frequency shifts.  The $\nu ST$ mechanism proposed above can be used to explain  both the glitch and ``anti-glitch'' phenomena. In particular, neutrinos (or antineutrinos) with a negative millicharged propagating inside the rotating star spin down the star rotation rate. From our estimations it follows that the proposed $\nu ST$ mechanism with about $10^{51}$ total emitted neutrinos and the neutrino millicharge $q_0=10^{-18}e_0$ can explain the recent ``anti-glitch'' event of the magnetar 1E 2259+586 \cite{Archibald:2013kla}.

\section{Conclusions}

In this work we have considered the electrically millicharged neutrino with the anomalous magnetic moment moving in the dense rotating magnetized matter. Two new exact solutions of the corresponding modified Dirac equations for the particle wave functions in the external environments have been obtained. It has been shown that the neutrino energy spectrum is quantized due to both weak and electromagnetic interactions of the neutrino with the background matter.

Within the quasi-classical interpretation of the obtained solutions we have introduced the effective Lorentz force that describes a millicharged neutrino motion in the rotating magnetized matter. Due to the action of the force low energy neutrinos can be bound in circular orbits inside neutron stars and accretion disks of black holes. In turn, the trajectories of the neutrinos with rather high energies will be reasonably disturbed. We have predicted the effect of the spatial separation of different types of neutrinos (neutrinos and antineutrinos, and/or different flavor neutrinos, and/or neutrinos with different energies) that occurs after passing of mixed flux of neutrinos through the rotating magnetized matter. This effect can explain the absence of correlations in observations of light and neutrino signals from their mutual astrophysical source in the ANTARES experiment~\cite{Ageron:2012mz}.

We have introduced a new possible mechanism of the neutrino electromagnetic radiation due to a nonzero neutrino millicharge (``Light of (milli)Charged Neutrino'', $LC\nu$) in the magnetized nonuniform rotating matter. This radiation can be produced when the neutrino is moving in the nonuniform rotating matter even in the absence of the magnetic field. The phenomenon is of interest for astrophysics in light of the recently reported hints of ultra-high energy neutrinos~$\sim$ PeV observed by IceCube~\cite{Abbasi:2012cu}.

We have also predicted the phenomenon of the star rotation frequency shift by neutrinos propagating inside a star (``Neutrino Star Turning" mechanism, $\nu S T$). The $\nu S T$ mechanism is originated due to both weak and electromagnetic interactions of the neutrinos with the dense magnetized matter of the star. The consideration of the impact of the proposed $\nu ST$ mechanism on the torque of pulsars during their formation in supernova explosions yields a new limit on the neutrino millicharge $q_{\nu}<1.3\times10^{-19} e_0$. The obtained limit is one of the strongest astrophysical bounds on the neutrino millicharge known in literature. In addition, the $\nu S T$ mechanism can be used to explain the origin of pulsar glitches and ``anti-glitches'' as well in case of a sufficiently large neutrino emission.

\section{Acknowledgments}
The authors are thankful to Sandip Pakvasa and Grisha Rubtsov for useful discussions.  The work on this paper has been partially supported by the Russian Foundation for Basic Research (grant No. 14-02-31816 mol\_a) and the Russian Science Foundation (grant No. 14-12-00033).





\bibliographystyle{model1a-num-names}
\bibliography{new_solution}

\begin{thebibliography}{40}
\expandafter\ifx\csname natexlab\endcsname\relax\def\natexlab#1{#1}\fi
\providecommand{\bibinfo}[2]{#2}
\ifx\xfnm\relax \def\xfnm[#1]{\unskip,\space#1}\fi
\bibitem[{Giunti and Studenikin(2009)}]{Giunti:2008ve}
\bibinfo{author}{C.~Giunti}, \bibinfo{author}{A.~Studenikin},
  \bibinfo{journal}{Phys. Atom. Nucl.} \bibinfo{volume}{72}
  (\bibinfo{year}{2009}) \bibinfo{pages}{2089}.
\bibitem[{Giunti and Studenikin(2014)}]{Giunti:2014ixa}
\bibinfo{author}{C.~Giunti}, \bibinfo{author}{A.~Studenikin},
  \bibinfo{journal}{arXiv:1403.6344}  (\bibinfo{year}{2014}).
\bibitem[{Broggini et~al.(2012)Broggini, Giunti, and
  Studenikin}]{Broggini:2012df}
\bibinfo{author}{C.~Broggini}, \bibinfo{author}{C.~Giunti},
  \bibinfo{author}{A.~Studenikin}, \bibinfo{journal}{Adv. High Energy Phys.}
  \bibinfo{volume}{2012} (\bibinfo{year}{2012}) \bibinfo{pages}{459526}.
\bibitem[{Studenikin(2009)}]{Studenikin:2008bd}
\bibinfo{author}{A.~Studenikin}, \bibinfo{journal}{Nucl. Phys. Proc. Suppl.}
  \bibinfo{volume}{188} (\bibinfo{year}{2009}) \bibinfo{pages}{220}.
\bibitem[{Fujikawa and Shrock(1980)}]{Fujikawa:1980yx}
\bibinfo{author}{K.~Fujikawa}, \bibinfo{author}{R.~Shrock},
  \bibinfo{journal}{Phys. Rev. Lett.} \bibinfo{volume}{45}
  (\bibinfo{year}{1980}) \bibinfo{pages}{963}.
\bibitem[{Beda et~al.(2012)}]{Beda:2012zz}
\bibinfo{author}{A.~Beda}, et~al., \bibinfo{journal}{Adv. High Energy Phys.}
  \bibinfo{volume}{2012} (\bibinfo{year}{2012}) \bibinfo{pages}{350150}.
\bibitem[{Bernstein et~al.(1963)Bernstein, Ruderman, and
  Feinberg}]{Bernstein:1963qh}
\bibinfo{author}{J.~Bernstein}, \bibinfo{author}{M.~Ruderman},
  \bibinfo{author}{G.~Feinberg}, \bibinfo{journal}{Phys. Rev.}
  \bibinfo{volume}{132} (\bibinfo{year}{1963}) \bibinfo{pages}{1227--1233}.
\bibitem[{Babu and Mohapatra(1990)}]{Babu:1989ex}
\bibinfo{author}{K.~S. Babu}, \bibinfo{author}{R.~N. Mohapatra},
  \bibinfo{journal}{Phys. Rev.} \bibinfo{volume}{D 41} (\bibinfo{year}{1990})
  \bibinfo{pages}{271}.
\bibitem[{Foot et~al.(1990)Foot, Joshi, Lew, and Volkas}]{Foot:1990uf}
\bibinfo{author}{R.~Foot}, \bibinfo{author}{G.~C. Joshi},
  \bibinfo{author}{H.~Lew}, \bibinfo{author}{R.~Volkas}, \bibinfo{journal}{Mod.
  Phys. Lett.} \bibinfo{volume}{A 5} (\bibinfo{year}{1990})
  \bibinfo{pages}{2721}.
\bibitem[{Foot et~al.(1993)Foot, Lew, and Volkas}]{Foot:1992ui}
\bibinfo{author}{R.~Foot}, \bibinfo{author}{H.~Lew},
  \bibinfo{author}{R.~Volkas}, \bibinfo{journal}{J. Phys} \bibinfo{volume}{G
  19} (\bibinfo{year}{1993}) \bibinfo{pages}{361}.
\bibitem[{Raffelt(1996)}]{Raffelt:1996wa}
\bibinfo{author}{G.~Raffelt}, \bibinfo{title}{Stars as laboratories for
  fundamental physics: The astrophysics of neutrinos, axions, and other weakly
  interacting particles}, \bibinfo{publisher}{University of Chicago Press},
  \bibinfo{year}{1996}.
\bibitem[{Barbiellini and Cocconi(1987)}]{Barbiellini:1987zz}
\bibinfo{author}{G.~Barbiellini}, \bibinfo{author}{G.~Cocconi},
  \bibinfo{journal}{Nature} \bibinfo{volume}{329} (\bibinfo{year}{1987})
  \bibinfo{pages}{21--22}.
\bibitem[{Marinelli and Morpurgo(1984)}]{Marinelli:1983nd}
\bibinfo{author}{M.~Marinelli}, \bibinfo{author}{G.~Morpurgo},
  \bibinfo{journal}{Phys. Lett.} \bibinfo{volume}{B 137} (\bibinfo{year}{1984})
  \bibinfo{pages}{439}.
\bibitem[{Baumann et~al.(1988)Baumann, Kalus, Gahler, and
  Mampe}]{Baumann:1988ue}
\bibinfo{author}{J.~Baumann}, \bibinfo{author}{J.~Kalus},
  \bibinfo{author}{R.~Gahler}, \bibinfo{author}{W.~Mampe},
  \bibinfo{journal}{Phys. Rev.} \bibinfo{volume}{D 37} (\bibinfo{year}{1988})
  \bibinfo{pages}{3107}.
\bibitem[{Studenikin and Ternov(2005)}]{Studenikin:2004dx}
\bibinfo{author}{A.~Studenikin}, \bibinfo{author}{A.~Ternov},
  \bibinfo{journal}{Phys. Lett.} \bibinfo{volume}{B 608} (\bibinfo{year}{2005})
  \bibinfo{pages}{107, hep--ph/0410296, hep--ph/0410297}.
\bibitem[{Grigorev et~al.(2005)Grigorev, Studenikin, and
  Ternov}]{Grigorev:2005sw}
\bibinfo{author}{A.~Grigorev}, \bibinfo{author}{A.~Studenikin},
  \bibinfo{author}{A.~Ternov}, \bibinfo{journal}{Phys. Lett.}
  \bibinfo{volume}{B 622} (\bibinfo{year}{2005}) \bibinfo{pages}{199}.
\bibitem[{Grigoriev et~al.(2007)Grigoriev, Savochkin, and
  Studenikin}]{Grigoriev:2007zzc}
\bibinfo{author}{A.~Grigoriev}, \bibinfo{author}{A.~Savochkin},
  \bibinfo{author}{A.~Studenikin}, \bibinfo{journal}{Russ. Phys. J.}
  \bibinfo{volume}{50} (\bibinfo{year}{2007}) \bibinfo{pages}{845}.
\bibitem[{Studenikin(2006{\natexlab{a}})}]{Studenikin20066769}
\bibinfo{author}{A.~Studenikin}, \bibinfo{journal}{J. Phys. A}
  \bibinfo{volume}{39} (\bibinfo{year}{2006}{\natexlab{a}})
  \bibinfo{pages}{6769--6776}.
\bibitem[{Studenikin(2006{\natexlab{b}})}]{Studenikin2006289}
\bibinfo{author}{A.~Studenikin}, \bibinfo{journal}{Ann. Found. de Broglie}
  \bibinfo{volume}{31} (\bibinfo{year}{2006}{\natexlab{b}})
  \bibinfo{pages}{289--316}.
\bibitem[{Studenikin(2008)}]{Studenikin:2008qk}
\bibinfo{author}{A.~Studenikin}, \bibinfo{journal}{J. Phys.} \bibinfo{volume}{A
  41} (\bibinfo{year}{2008}) \bibinfo{pages}{164047}.
\bibitem[{Balantsev et~al.(2011)Balantsev, Popov, and
  Studenikin}]{Balantsev:2010zw}
\bibinfo{author}{I.~Balantsev}, \bibinfo{author}{Y.~Popov},
  \bibinfo{author}{A.~Studenikin}, \bibinfo{journal}{J. Phys}
  \bibinfo{volume}{A 44} (\bibinfo{year}{2011}) \bibinfo{pages}{255301}.
\bibitem[{Ageron et~al.(2012)}]{Ageron:2012mz}
\bibinfo{author}{M.~Ageron}, et~al., \bibinfo{journal}{Nucl. Instrum. Meth.}
  \bibinfo{volume}{A 692} (\bibinfo{year}{2012}) \bibinfo{pages}{184--187}.
\bibitem[{Abbasi et~al.(2012)}]{Abbasi:2012cu}
\bibinfo{author}{R.~Abbasi}, et~al., \bibinfo{journal}{Phys. Rev.}
  \bibinfo{volume}{D 86} (\bibinfo{year}{2012}) \bibinfo{pages}{022005}.
\bibitem[{Archibald et~al.(2013)}]{Archibald:2013kla}
\bibinfo{author}{R.~Archibald}, et~al., \bibinfo{journal}{Nature}
  \bibinfo{volume}{497} (\bibinfo{year}{2013}) \bibinfo{pages}{591--593}.
\bibitem[{Studenikin and Tokarev(2013)}]{Studenikin:2013yaa}
\bibinfo{author}{A.~Studenikin}, \bibinfo{author}{I.~Tokarev},
  \bibinfo{journal}{Nucl.Phys.Proc.Suppl.} \bibinfo{volume}{237-238}
  (\bibinfo{year}{2013}) \bibinfo{pages}{317--319}.
\bibitem[{Balantsev et~al.(2012)Balantsev, Studenikin, and
  Tokarev}]{Balantsev:2012ep}
\bibinfo{author}{I.~Balantsev}, \bibinfo{author}{A.~Studenikin},
  \bibinfo{author}{I.~Tokarev}, \bibinfo{journal}{Phys. Part. Nucl.}
  \bibinfo{volume}{43} (\bibinfo{year}{2012}) \bibinfo{pages}{727--741}.
\bibitem[{Balantsev et~al.(2013)Balantsev, Studenikin, and
  Tokarev}]{Balantsev:2013aya}
\bibinfo{author}{I.~Balantsev}, \bibinfo{author}{A.~Studenikin},
  \bibinfo{author}{I.~Tokarev}, \bibinfo{journal}{Phys. Atom. Nucl.}
  \bibinfo{volume}{76} (\bibinfo{year}{2013}) \bibinfo{pages}{489--503}.
\bibitem[{Frolov and Zhukovsky(2007)}]{Frolov:2007cm}
\bibinfo{author}{I.~Frolov}, \bibinfo{author}{V.~C. Zhukovsky},
  \bibinfo{journal}{J. Phys.} \bibinfo{volume}{A 40} (\bibinfo{year}{2007})
  \bibinfo{pages}{10625--10640}.
\bibitem[{Oliveira~e Silva et~al.(2000)}]{OliveiraeSilva:2000ns}
\bibinfo{author}{L.~Oliveira~e Silva}, et~al., \bibinfo{journal}{Phys. Lett.}
  \bibinfo{volume}{A 270} (\bibinfo{year}{2000}) \bibinfo{pages}{265}.
\bibitem[{Loeb(1990)}]{Loeb:1989nb}
\bibinfo{author}{A.~Loeb}, \bibinfo{journal}{Phys. Rev. Lett.}
  \bibinfo{volume}{64} (\bibinfo{year}{1990}) \bibinfo{pages}{115}.
\bibitem[{Schwinger(1949)}]{Schwinger:1949ym}
\bibinfo{author}{J.~S. Schwinger}, \bibinfo{journal}{Phys. Rev.}
  \bibinfo{volume}{75} (\bibinfo{year}{1949}) \bibinfo{pages}{1912}.
\bibitem[{Oraevsky et~al.(1994)Oraevsky, Semikoz, and
  Smorodinsky}]{Oraevsky:1994wb}
\bibinfo{author}{V.~Oraevsky}, \bibinfo{author}{V.~Semikoz},
  \bibinfo{author}{Y.~Smorodinsky}, \bibinfo{journal}{Phys. Part. Nucl.}
  \bibinfo{volume}{25} (\bibinfo{year}{1994}) \bibinfo{pages}{129--156}.
\bibitem[{Nieves(2003)}]{Nieves:2003kw}
\bibinfo{author}{J.~F. Nieves}, \bibinfo{journal}{Phys. Rev.}
  \bibinfo{volume}{D 68} (\bibinfo{year}{2003}) \bibinfo{pages}{113003}.
\bibitem[{Duan and Qian(2004)}]{Duan:2004nc}
\bibinfo{author}{H.~Duan}, \bibinfo{author}{Y.-Z. Qian},
  \bibinfo{journal}{Phys. Rev.} \bibinfo{volume}{D 69} (\bibinfo{year}{2004})
  \bibinfo{pages}{123004}.
\bibitem[{Mikaelian(1977)}]{Mikaelian:1976ce}
\bibinfo{author}{K.~O. Mikaelian}, \bibinfo{journal}{Astrophys.J.}
  \bibinfo{volume}{214} (\bibinfo{year}{1977}) \bibinfo{pages}{L23}.
\bibitem[{Epstein(1978)}]{Epstein:1978md}
\bibinfo{author}{R.~Epstein}, \bibinfo{journal}{Astrophys.J.}
  \bibinfo{volume}{219} (\bibinfo{year}{1978}) \bibinfo{pages}{L39--L41}.
\bibitem[{Hirata et~al.(1987)}]{Hirata:1987hu}
\bibinfo{author}{K.~Hirata}, et~al., \bibinfo{journal}{Phys. Rev. Lett.}
  \bibinfo{volume}{58} (\bibinfo{year}{1987}) \bibinfo{pages}{1490--1493}.
\bibitem[{Rosen et~al.(2013)}]{Rosen:2012ty}
\bibinfo{author}{R.~Rosen}, et~al., \bibinfo{journal}{Astrophys. J.}
  \bibinfo{volume}{768} (\bibinfo{year}{2013}) \bibinfo{pages}{85}.
\bibitem[{Howell et~al.(2013)}]{Howell:2013qua}
\bibinfo{author}{D.~Howell}, et~al., \bibinfo{journal}{Astrophys.J.}
  \bibinfo{volume}{779} (\bibinfo{year}{2013}) \bibinfo{pages}{98}.
\bibitem[{Anderson and Itoh(1975)}]{Anderson:1975zze}
\bibinfo{author}{P.~Anderson}, \bibinfo{author}{N.~Itoh},
  \bibinfo{journal}{Nature} \bibinfo{volume}{256} (\bibinfo{year}{1975})
  \bibinfo{pages}{25--27}.

\end{thebibliography}







\end{document}